\documentclass[prb, A4paper, twocolumn, amsfonts, superscriptaddress, floatfix]{revtex4-2}

\usepackage[utf8]{inputenc}
\usepackage{amsmath, amssymb, amsfonts}
\usepackage{graphicx, color, bm, float}
\usepackage{mathptmx}
\usepackage{tikz, tikzscale, svg}
\usepackage{fancyhdr}
\usepackage{lastpage}

\DeclareRobustCommand\dotted{\tikz[baseline=-0.6ex]\draw[thick, line cap=round, dash pattern= on 1pt off 5pt] (0,0)--(0.48,0);}
\DeclareRobustCommand\dashed{\tikz[baseline=-0.6ex]\draw[thick, line cap=round, dash pattern= on 3pt off 2pt] (0,0)--(0.48,0);}
\DeclareRobustCommand\chain {\tikz[baseline=-0.6ex]\draw[thick, line cap=round, dash pattern= on 2pt off 2pt on 6pt off 2pt] (0,0)--(0.48,0);}

\newcommand*{\bluecolor}[1]{\textcolor{blue}{#1}}

\usepackage[pdfstartview = FitV,
            bookmarks    = true,
            linktocpage  = true,
            colorlinks   = true,
            linkcolor    = blue,
            urlcolor     = blue,
            citecolor    = green,
            anchorcolor  = blue]{hyperref}     
            
\hypersetup{
    pdfauthor={A. Charau, J. Laurent, and T. Valier-Brasier},
    pdfcreator={A. Charau, J. Laurent, and T. Valier-Brasier},
    pdfproducer={alexandre.charau@cea.fr et al.},   
    pdftitle={Adiabatic Lamb modes in 3D tapered waveguides: Cut-off effects and ZGV resonances},
    pdfsubject={CEA and IJRA, (2025)},
    pdfkeywords = {Adiabatic modes, Inhomogeneous waveguide, Laser-Ultrasound, 2D reconstruction of thickness profile, cut-off and ZGV thicknesses}}            
\graphicspath{{./fig/}}
\makeatletter
\def\section{\@startsection
  {section}{1}{\z@}%
  {-2ex \@plus -0.3ex \@minus -.1ex}
  {2ex \@plus .3ex}
  {\centering\normalfont\bfseries}}
\makeatother

\everymath{\displaystyle}

\begin{document}

\makeatletter
\let\ps@titlepage\ps@fancy
\makeatother

\title[Draft Ultrasonics]{Adiabatic Lamb modes in 3D tapered waveguides: \\Cut-off effects and ZGV resonances}

\author{Alexandre Yoshitaka Charau}
\email{alexandre.charau@cea.fr}
\affiliation{\href{https://ror.org/03xjwb503}{Universit\'e Paris-Saclay}, \href{https://ror.org/000dbcc61}{CEA, List}, F-91120, Palaiseau, France}
\affiliation{\href{https://ror.org/043we9s22}{Institut Jean Le Rond $\partial$'Alembert}, \href{https://ror.org/02en5vm52}{Sorbonne Universit\'e}, UMR CNRS7190, 4 Place Jussieu, Paris 75005, France\vskip 3mm\footnotesize{(Received 26 February 2025; Received in revised form 28 May 2025; Accepted 14 June 2025)}}

\author{J\'er\^ome Laurent}
\email{jerome.laurent2@cea.fr}
\affiliation{\href{https://ror.org/03xjwb503}{Universit\'e Paris-Saclay}, \href{https://ror.org/000dbcc61}{CEA, List}, F-91120, Palaiseau, France}

\author{Tony Valier-Brasier}
\email{tony.valier-brasier@sorbonne-universite.fr}
\affiliation{\href{https://ror.org/043we9s22}{Institut Jean Le Rond $\partial$'Alembert}, \href{https://ror.org/02en5vm52}{Sorbonne Universit\'e}, UMR CNRS7190, 4 Place Jussieu, Paris 75005, France\vskip 3mm\footnotesize{(Received 26 February 2025; Received in revised form 28 May 2025; Accepted 14 June 2025)}}


\begin{abstract}
This paper aims to enhance our understanding of the physical behavior of adiabatic modes in inhomogeneous elastic plates, particularly their remarkable capacity to adapt to gradual perturbations. The study investigates the propagation characteristics of higher-order adiabatic Lamb modes in waveguides with linearly varying thickness, with a focus on the influence of critical thicknesses on their propagation. This is achieved by leveraging the broadband excitation capabilities of a pulsed laser generating higher order Lamb modes to reveal various critical thicknesses, such as the cut-off and Zero-Group Velocity (ZGV) thicknesses. Remarkably, ZGV resonances can be induced at locations well beyond the laser source. Moreover, the mode's behavior is strongly influenced by thickness variations in all directions, imparting the plate an anisotropic-like behavior. Additionally, based on the observed effects, our experimental approach enables precise reconstruction of elastic waveguide profiles in additively manufactured aluminum plates with such thickness variations. The reconstructed profiles show a strong correlation with reference measurements across the scanned area.
\vskip 2mm
\noindent DOI: \href{https://doi.org/10.1016/j.ultras.2025.107733}{\nolinkurl{10.1016/j.ultras.2025.107733}}
\end{abstract}

\maketitle
\newpage

\section{Introduction}\label{Introduction}

A substantial body of research has been conducted on guided modes in free elastic plates with smoothly varying thickness, employing a range of approaches, including theoretical, numerical, and experimental techniques~\citep{pagneux_lamb_2006, postnova_trapped_2007, Valier2008,  zima_theoretical_2022}. These studies have considerably enhanced our comprehension of guided mode propagation in such inhomogeneous and tapered waveguides~\cite{biryukov1995}, employing observed phenomena to more accurately characterize these waveguides. In regions where the thickness changes gradually, Lamb modes exhibit subtle modifications~\cite{lamb_waves_1917}, including the generation of adiabatic modes~\cite{pierce_extension_1965, santini_elastic_2023}. The analogy between adiabatic modes in elastic waveguides and the adiabatic approximation in quantum mechanics is based on the unchanging eigenstate of a system (or mode) as it adapts instantaneously and locally to slow variations in its conditions or graded parameters (Born-Oppenheimer approximation)~\citep{born_zur_1927, born_beweis_1928}. The investigation of adiabatic modes commenced with efforts to understand the propagation of normal modes in acoustic waveguides, especially within stratified oceanic environments. Initially, this research was highly theoretical and led to the development of various analytical models predicting the behavior of these distinctive modes~\citep{pierce_guided_1982, arnold_local_1986}. Seminal studies have demonstrated the presence of adiabatic modes in elastic waveguides through both numerical simulations and experimental observations~\citep{el-kettani_guided_2004, marical_guided_2007}. Similar observations were made shortly afterwards with EMAT transducers using the SH or torsional guided waves~\citep{Nakamura2012, Nakamura2013}, characterized by the sudden decrease in traveling time after the cut-off thickness.\\

Adiabatic modes adapt to variations in waveguide thickness by continuously adjusting their phase velocity $V_{\varphi}$ and therefore their wavenumber $k$ during propagation, mirroring the local thickness at each point in the medium. Consequently, for relatively short propagation distances, a waveguide with varying thickness can be approximated as a series of segments, each with a uniform thickness. As the thickness decreases, the higher-order adiabatic mode approaches its critical thickness $h_c$, resulting in a wave reflection within the plate~\citep{arnold_intrinsic_1984, felsen_adiabatic_1991, rose_using_1998, belanger_high_2014, suresh_remnant_2022}. A recent analytical study has investigated the potential for reconstructing waveguide shapes with gradual thickness variations by exploiting the thickness resonances~\cite{niclas_reconstruction_2023}. The study explores how these resonances can be used to achieve detailed reconstruction of geometrical shapes by solving the inverse problem with the associated Airy function. This approach is particularly significant when a higher-order adiabatic mode propagates in a waveguide with a linearly decreasing thickness, as it may locally encounter its critical thickness~\cite{hamitouche_reflection_2009}. When these critical thicknesses correspond to cut-off conditions~\citep{arnold_intrinsic_1984, felsen_adiabatic_1991, marical_guided_2007}, the real and imaginary parts of the wavenumber $k$ are both equal to zero, which implies that $V_{\varphi}=2\pi f/\Re(k)$ approaches infinity. An infinite phase velocity indicates that the wavefront or phase advances at an unlimited phase velocity, which implies that there is no actual physical displacement of ultrasound within the medium. When the frequency diminishes below the mode's cut-off frequency, the wavenumber becomes imaginary, denoting the transition to evanescent mode, which are preventing mode propagation and resulting in an exponential attenuation of its amplitude. This transition is consistent with the behavior of adiabatic modes in such waveguides. During the transition at the turning points, where oscillatory solutions connect to evanescent ones, mode conversions may occur, enabling lower-order modes to propagate beyond the cut-off thickness~\citep{arnold_intrinsic_1984, felsen_adiabatic_1991,rose_using_1998, yan_conversion_2015}. This phenomenon, which is sometimes referred to as the ``tunneling effect'', is supported by previous research~\citep{arnold_local_1986, hamitouche_reflection_2009, germano_tunnel_2024}. At the cut-off frequency, thickness resonance occurs, and like all guided modes, the mode has both in-plane and out-of-plane displacement components. The sensitivity to this resonance may vary depending on which component is being measured.\\

Therefore, cut-off thicknesses are not the only thicknesses at which a mode resonates or cease to propagate. ZGV modes occur at distinct points on the dispersion curves where the group velocity,  $V_g=\partial \omega / \partial k$ of the wave is zero, while the phase velocity remains finite \cite{prada_laser-based_2005}. This unique behavior renders ZGV modes exceptionally sensitive to local variations in material properties or thickness, making them ideal for high-resolution material characterization \cite{ces_thin_2011, yan2018, legrand2018, kiefer2023, thelen2021,  morales_acoustoelastic_2024}. Unlike cut-off thicknesses, which delineate the transition between propagating and non-propagating states, ZGV modes are distinguished by localized resonances that trap energy at the excitation point, rather than allowing it to propagate. A defining feature of ZGV modes is that their wavenumbers correspond to the saddle points on the dispersion curves of the mode to which they belong. Astonishingly, we show that ZGV resonances can also occur at points far away from the laser source. By leveraging this phenomenon in a waveguide having a linearly varying thickness, these remarkable points can also naturally be reached due to the adiabatic regime.

Building on established research and physical principles, this paper presents two experimental methods that leverage the adiabatic principle. The first method focuses on the cut-off thicknesses of the first-order antisymmetric Lamb mode ($A_1$), while the second exploits the first Zero-Group Velocity (ZGV) thicknesses ($S_1S_2$), both induced by laser excitation~\cite{scruby_laser_1990}. These techniques enable the precise reconstruction of a metallic waveguide profile, fabricated via additive manufacturing (AM) and characterized by linear thickness variations in all directions.

\section{Experimental setup}\label{Expe}

The experimental setup is depicted in Fig.~(\ref{Expe}). The experiments were conducted on an aluminium plate (AlSi10Mg) measuring 250$\times$250 mm$^2$, produced using the Laser Powder Bed Fusion (LPBF) process. The plate exhibits a series of linear thickness variations, with corner thicknesses of 5, 4, 3 and 2 mm, respectively. The thickness variation is present on only one side of the plate. X-ray radiography also confirmed that the AM-plate is free of internal defects, such as cracks or porosities (details are given in the Supplementary Material~\bluecolor{S1}). Lamb modes were generated using a Q-switched Nd:YAG laser (Quantum Light, Q2HE) with a wavelength of 1064 nm. The laser emitted pulses with a duration of 7 ns and the energy has adjusted to 2 mJ (thermoelastic regime) with a repetition rate of 60 Hz. The galvanometer head (Thorlabs GVS012/M) and the F-$\theta$ lens (Thorlabs FTH254-1064), positioned at the laser output, facilitated precise scanning of the target scanned area, producing a focal spot size of 1 mm. 

\begin{figure*}[ht!]
\centering
    \includegraphics[width=0.9\linewidth]{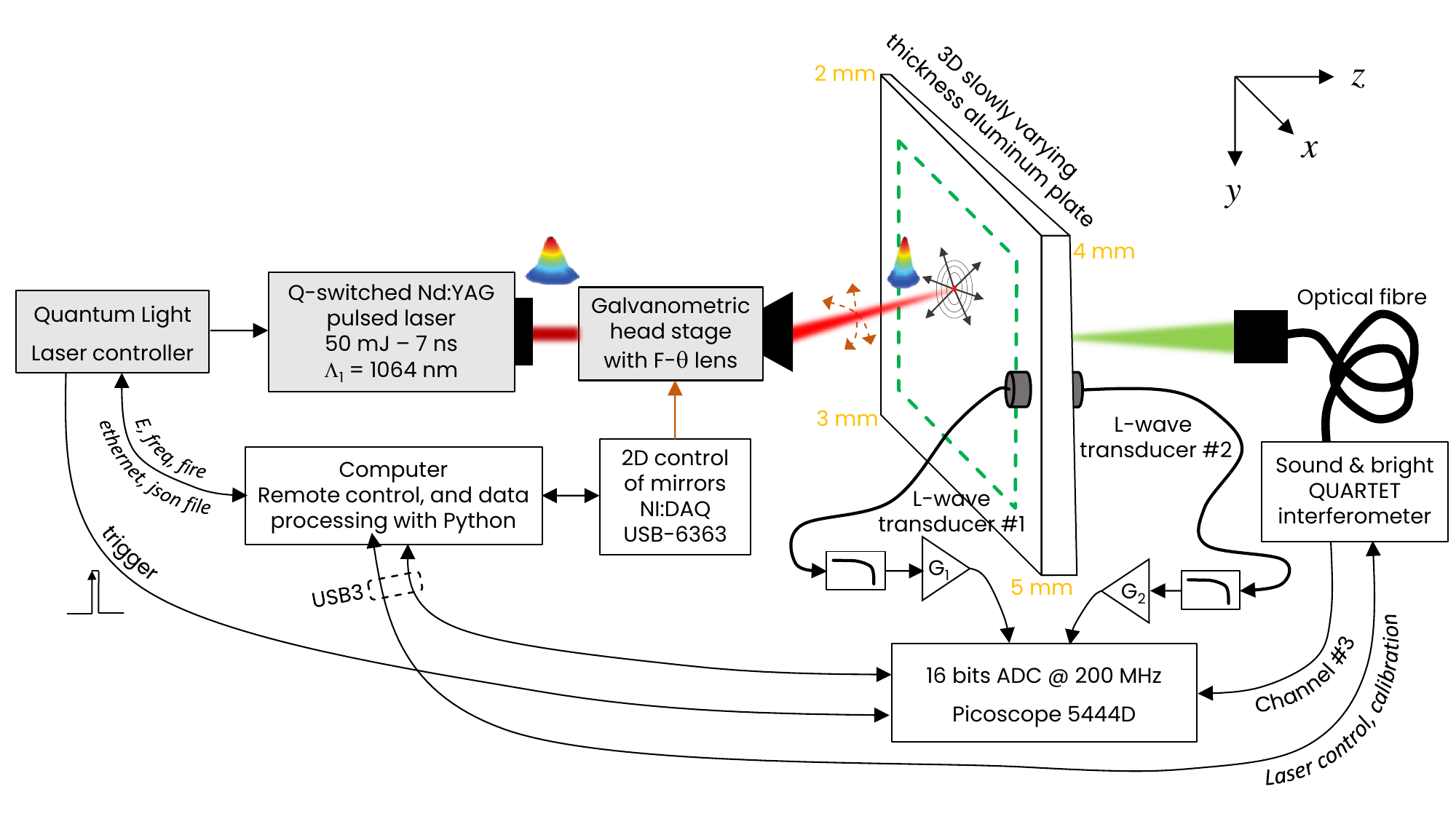}
    \caption{Experimental setup for the laser generation and probing of Lamb modes on an elastic AM-plate with a slow variation of the thickness in all directions.This experimental setup is based on the principle of reciprocity: the pulsed laser is in motion and both longitudinal transducers is in the probing position to obtain a better signal-to-noise ratio.}
  \label{Expe}
\end{figure*}

The scanning system is controlled by a DAQ card (USB-6363, National Instruments) with a dynamic range of up to 16 bits, enabling highly precise control of the mirrors. The normal displacement wavefield was measured using a Krautkramers contact transducers (model 015PV1) with a nominal frequency of 1 MHz, leveraging the principle of reciprocity \cite{achenbach2004}. The signal from both transducers is processed through a low-pass filter with a cut-off frequency of 10 MHz and then amplified using two low-noise preamplifiers from Panametrics (5077PR), which has a maximum gain of 40 dB ($G_{1,2}$), before being digitized. The Quartet interferometer was employed to measure the material properties of the plate at a number of points through the first two ZGV resonances (see the Supplementary Material~\bluecolor{S2}). A single transducer is used to analyze the $A_1$-cut-off thicknesses. However, in the case of the ZGV thicknesses, both transducers are employed, a choice that is justfied in detail in a section~(\ref{ZGV_thickness_sec}) to favor the selective detection of symmetric modes. Data collection was performed with a Picoscope card (5444D, from PicoTech), offering a dynamic resolution of 16 bits. The scanning system covered a 120 x 120 mm$^2$ area with a step size of 0.4 mm, a sampling frequency of 40 MHz, and an observation time set at 200 µs per scan. The entire scanning process took approximately six hours. Furthermore, a CMOS type micro laser distance sensor with 30 µm repeatability (Panasonic HG-C1050) mounted on a Universal Robot (UR5) was employed with the picoscope card to conduct a secondary scan over the identical scanned ROI, thereby providing a reference thickness map. This inhomogeneous elastic plate meets the adiabatic conditions required for the A${_1}$-mode propagation. For further details, please refer to the adiabaticity criteria outlined in reference~\cite{moreau_measuring_2014} and further details can be found in the Supplementary Material~\bluecolor{S3}. 

To estimate the critical thicknesses for the studied inhomogeneous plate, it is necessary to know the theoretical dispersion curves for the samples. The dispersion curves are obtained by solving the Rayleigh-Lamb equation~\cite{royer_ondes_2021}, using the longitudinal ($V_L$) and transverse ($V_T$) bulk wave velocities, which are experimentally determined from the first two Zero-Group-Velocity (ZGV) resonances~\citep{clorennec_local_2007, grunsteidl_determination_2018}. The measurement process was conducted using a plate of the same material but with a uniform thickness. The same pulsed laser and Quartet laser interferometer from Sound\&Bright were used to measure the normal displacement at a few specific points on the plate. The resulting bulk wave velocities are: V$_L$ = 6.32 µs.mm$^{-1}$ and V$_T$ = 3.12 µs.mm$^{-1}$ ($\nu = 0.34$). This method, which relies on ZGV modes, provides a localized measurement of the plate's material properties, in contrast to the global approach that will be presented later in this study.

\section{$A_1$-mode cut-off thicknesses}\label{A1cut-off}

\begin{figure}[!ht]
\centering
    \includegraphics[width=0.9\columnwidth]{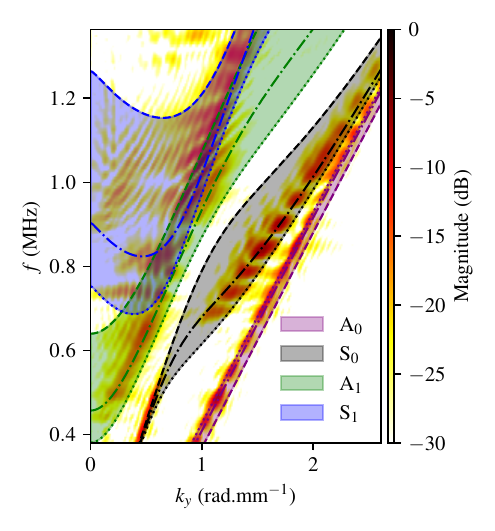}
    \caption{Experimental dispersion curves, measured with transducer \#1, for the column at $x = 60$ mm are compared with theoretical curves calculated for the following thicknesses: $h_1$ = 4.2 mm (\dotted), $h_m$ = 3.5 mm (\chain), and $h_2$ = 2.5 mm (\dashed).}
    \label{2D_disp}
\end{figure}

Figure~(\ref{2D_disp}) shows both the theoretical and experimental dispersion curves measured with the transducer \#1. The theoretical curve is obtained semi-analytically using a in-house Python code, while the experimental curve is derived from the 2D scan described earlier, using a 2D-FFT applied to the central column at $x=60$ mm. In this scan, the waveguide thickness is observed to vary from a maximum value of $h_1$ = 4.2 mm to a minimum value of $h_2$ = 2.5 mm. Unlike dispersion curves for a waveguide with a constant cross-section, the energy of the modes spreads out for the studied waveguide. This spreading occurs due to the change in wavelength experienced by the guided modes, which is directly caused by the gradual change in thickness. However, a comparison with the theoretical dispersion curves reveals that the energy is not distributed uniformly across the spatial bandwidth of the modes for a given frequency. Instead, it is predominantly concentrated around the dispersion curves for a plate with an effective thickness of $h_m=3.5$ mm, which is the average of $h_1$ and $h_2$. 

This behavior is directly related to the energy conservation properties of adiabatic modes, which are influenced by modal amplitudes (results of the excitability calculation are given in the Supplementary Material~\bluecolor{S4}). As propagation occurs towards thinner sections of the waveguide, the energy density of the adiabatic mode increases in the thickness direction. Simultaneously, the out-of-plane excitability of the modes influences the modal amplitude of a given mode for a specific wavenumber, frequency, and thickness (see Supplementary Material \bluecolor{S5}). As previously indicated, the plate exhibits variations in thickness across all dimensions. To elucidate the propagation and identify the cut-off frequencies of the $A_1$-mode in various directions, it is crucial to consider these variations. In particular, an examination of how the bandwidth changes with wavenumber in different directions provides insights into the anisotropic effects caused by the non-uniform thickness of the specimen. In order to map these bandwidth variations in a systematic manner, it is necessary to conduct a comprehensive analysis of the wavenumber. This enhances our understanding of the dispersion behavior and establishes critical thickness thresholds for $A_1$-mode propagation in multiple orientations.

\begin{figure*}[ht!]
\centering
    \includegraphics[width=0.8\linewidth]{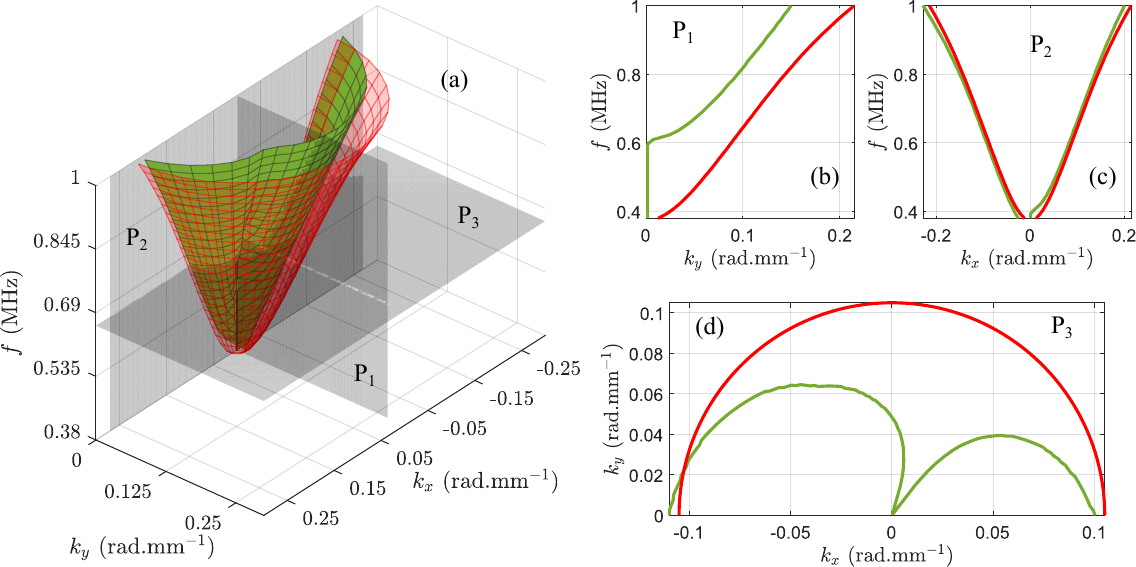}
    \caption{Theoretical parametric dispersion curves for the $A_1$-mode: (a) for a constant thickness $h_1$ (red) and for observed thicknesses at the edge of the scanned zone $h_e$ (green); projection of the dispersion curves presented in three different $(f,k_x,k_y)$ planes for a better overview: (b) $P_1 \to k_x=0$, (c) $P_2 \to k_y=0$, (d) and $P_3 \to f=0.5$ MHz.}
  \label{3D_disp}
\end{figure*}

In order to achieve this, we compute two sets of three-dimensional theoretical dispersion curves calculated for the incident $A_1$-mode for each direction. The initial set of theoretical dispersion curves is based on a constant thickness $h_1$, which is measured at the reference point ($x$ = 60 mm, $y$ = 0 mm). The second set of curves matches to the thickness observed at the edge of the scanned area $h_e$. Figure~(\ref{3D_disp}) illustrates the variations in wavenumber bandwidth of the incident $A_1$-mode across different directions. It is noteworthy that the Fig.~(\ref{3D_disp}d) demonstrates the anisotropic-like nature of this plate, indicating that at 0.5 MHz, the $A_1$-mode reaches its cut-off frequency, which varies with the direction. 

\begin{figure*}[ht!]
    \centering
    \includegraphics[width=0.9\linewidth]{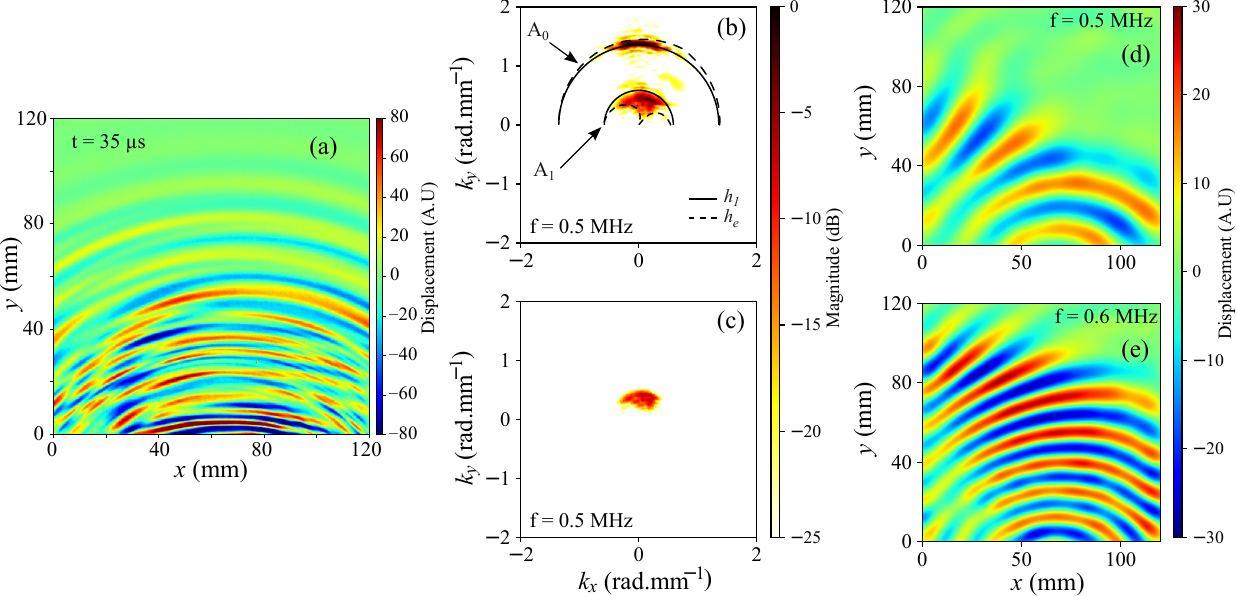}
    \caption{(a) Experimental wavefield at 35 µs propagation time; (b) 2D spatial Fourier plane at 0.5 MHz ; (c) 2D filtered spatial Fourier plane at 0.5 MHz ;  Wavefield (real part) of incident $A_1$-mode at 0.5 MHz (d) and 0.6 MHz (e) obtained after spatial IFFT.}
    \label{exp_scan}
\end{figure*}

This methodology enables the observation of the propagation of the $A_1$-mode at a specific frequency, propagating through a plate with a linearly varying thickness. This allows for the identification of the transition from propagative to evanescent modes. Consequently, it becomes possible to experimentally determine the location where propagation fades, which correlates to the local cut-off thickness of the plate. The observed wavefield in Fig.~(\ref{exp_scan}a) illustrates the significant difficulty of identifying and isolating the distinct guided modes, which is attributed to the broadband nature of the excitation induced by the pulsed laser and measured by the transducer. This challenge is compounded by the multi-directional linear variations in the plate, which introduce a phase and group velocity gradient in each direction, making it behave somewhat like an anisotropic waveguide. This observation is consistent with the theoretical phenomenon of anisotropy discussed earlier. A three-dimensional Fast Fourier Transform (3D FFT) is performed, thereby facilitating access to the three-dimensional experimental dispersion curves. Figure~(\ref{exp_scan}b), depicts the experimental spatial Fourier plane, overlaid with the theoretical curves of A$_0$-mode and $A_1$-mode shown in Fig.~\ref{3D_disp}(d) at $f=0.5$ MHz.

There is a clear agreement between the theoretical and experimental curves, which serves to further demonstrate the anisotropic-like nature when observed in the Fourier domain. It can be seen that the visible energy between the A$_0$-mode and $A_1$-mode corresponds to the S$_0$-mode; however, for the sake of clarity, the curves for this mode are not displayed. In order to isolate the contribution of the A$1$-mode from the wavefield at each frequency $f$, a Tukey window (shape parameter $\alpha = 0.9$) filter is applied within the spatial Fourier plane. The filter permits the extraction of information pertaining to the $A_1$-mode at positive wavenumbers $k_x$ and $k_y$ spanning a range from 0 rad.mm$^{-1}$ to $k_{h_{max}}$ [Fig.~(\ref{exp_scan}c)]. In this context, $k_{h_{max}}$ represents the maximum wavenumber for the $A_1$-mode, corresponding to the thickest cross-section of the waveguide. By employing such filter within these values, only the incident propagating modes are captured. Subsequently, an inverse FFT is applied in order to capture the wavefield corresponding to the $A_1$-mode at the targeted frequency. Figures~(\ref{exp_scan}d) and~(\ref{exp_scan}e) show the $A_1$-mode wavefield at 0.5 and 0.6 MHz, respectively. As the frequency increases, the mode propagates more prominently in the thinner sections of the guide. This phenomenon demonstrates that at a specific frequency, the wavefield fades out at the cut-off thickness.

\begin{figure*}[ht!]
    \centering
    \includegraphics[width=0.9\linewidth]{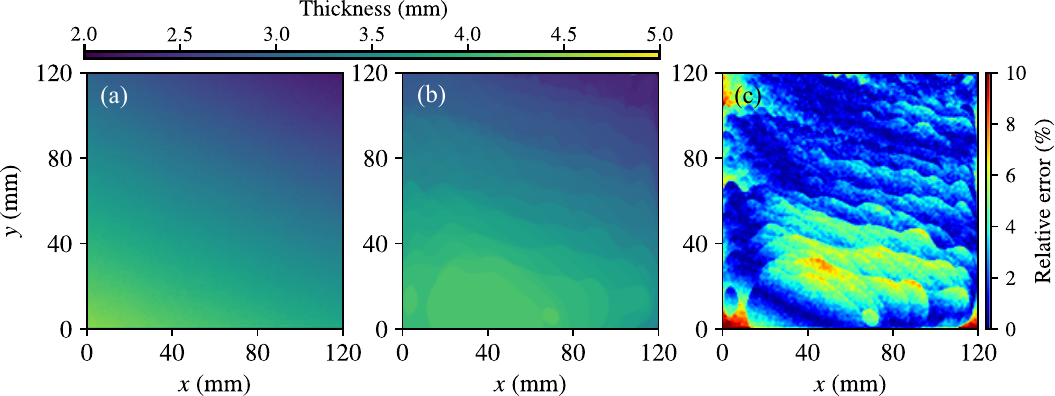}
    \caption{(a) Measured thickness map with the laser rangefinder, (b) reconstructed thickness map, and (c) relative error between thickness maps.}
    \label{final_res}
\end{figure*}

As $(\omega,k_x,k_y)$-plane processing was applied across the entire frequency range, it is possible to visualize the $A_1$-mode wavefield at any frequency. The wavefields have been reconstructed for a frequency range from 0.38 MHz to 0.8 MHz, which corresponds to the minimum and maximum cut-off frequencies for the $A_1$-mode in the waveguide, respectively. As a consequence of the inverse FFT, the filtered field is complex. In order to facilitate the observation of the mode halting at its cut-off thickness, the absolute value of the wavefields is used. On the basis of the $A_1$-mode dispersion curves, modal amplitudes, and in particular its cut-off frequencies for different thicknesses, it can be inferred that areas with detectable energy are those where the mode is propagative. This suggests that the material thickness in these regions meets or exceeds the cut-off thickness. Furthermore, the $A_1$-mode's excitability at varying frequencies was taken into account during the thresholding process. Ultimately, by employing an iterative approach to analyze the data from the highest to the lowest frequency, we are able to reconstruct the waveguide thickness. Fig.~(\ref{final_res}a) depicts the thickness map generated using the aforementioned method, exhibiting a discernible trend of decreasing thickness as one progresses towards higher values of the $x$ and $y$ coordinates. The layered appearance observed in Fig.~(\ref{final_res}a) is a consequence of the frequency step. 

A more precise reconstruction is likely to be achieved by employing a finer frequency step, either by extending the acquisition time or by utilising zero-padding. The degree of error increases with increasing thickness, due to the necessity of lower frequency observations for the accurate reconstruction of these thicker sections. As the transducer is centered at 1 MHz, it is not well-suited for capturing the necessary low-frequency data, which limits the precision of the reconstruction. Furthermore, errors may result from the filtering process applied in the spatial Fourier domain. Despite these challenges, the reconstructed image exhibits a high degree of qualitative similarity to the reference thickness map, as illustrated in Fig.~(\ref{final_res}b). In terms of quantitative comparison, Fig.~(\ref{final_res}c) depicts the relative error between the thickness map reconstructed with our method and the one obtained using the laser rangefinder, with an average relative error close to 2.6\%.

\section{$S_1S_2$-ZGV mode thicknesses}\label{ZGV_thickness_sec}

In this section, we harness the surprising potential to generate ZGV resonances at any desired location in the plate, even those far from the laser source, in order to reconstruct the smooth thickness profile more accurately. To enhance the signal-to-noise ratio (SNR), we exploit both the principle of reciprocity and the subtraction of signals from the two longitudinal transducers, thereby refining the selective probing of symmetric modes and removing the limitation imposed by the $A_1$-mode, which would otherwise truncate the $S_1S_2$-branch.

\begin{figure*}[ht!]
    \centering
    \includegraphics[width=0.8\linewidth]{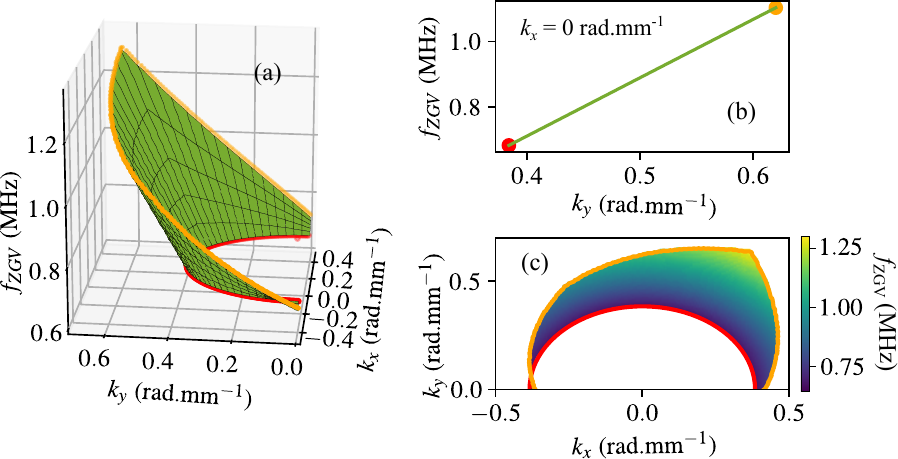}
    \caption{Theoretical parametric dispersion curves for the $S_1S_2$-ZGV mode are presented for two scenarios: a constant thickness $h_1$ (red) and the observed thicknesses at the edge of the scanned zone $h_e$ (orange) (a). The dispersion curves are superimposed onto two different $(f,k_x,k_y)$ planes to facilitate visualization: at $k_x=0$ (b) and $(k_x,k_y)$ plane with the frequency in color scale (c).}
    \label{ZGV_existency}
\end{figure*}

In contrast to the analysis of modes over a range of bandwidths, as conducted for dispersion curves of waveguides with variable cross-sections, the investigation of ZGV modes focuses on remarkable points on the dispersion curve where the group velocity tends to zero~\cite{royer_elastic_2022,prada_laser-based_2005}. This methodology is fundamentally distinct in that it does not necessitates the examination of a continuous range of wavenumbers from $k_{min} = 0$ rad.mm$^{-1}$ to a finite $k_{max}$. Instead, it is focused on the identification of the resonance that corresponds to the ZGV condition. Figure~(\ref{ZGV_existency}) illustrates the theoretical locations within the scanned region of interest where $S_1S_2$-ZGV modes exist. This representation clearly reflects the effects associated with geometric anisotropy-like, which are typically linked to the varying thickness of the waveguide. The localization of the $S_1S_2$-ZGV modes is governed by the interplay between the waveguide's geometry and its dispersion characteristics, which demonstrates that the ZGV condition is satisfied. These points correspond to specific thicknesses in the scanned area, which are directly related to the ZGV frequencies. This mapping provides a clear visualization of the localized nature of ZGV mode energy, which is distinct from the broader energy distribution described in the analysis presented earlier. Figure~(\ref{ZGV_existency}b) illustrates that the relationship between the $S_1S_2$-ZGV frequency and the wavenumber is linear, which can be attributed to the resonance condition of the ZGV mode itself being linearly dependent on the waveguide thickness. This dependence indicates that the spatial extent of the $S_1S_2$-ZGV mode varies according to the reconstructed zone. In particular, the width of the reconstructed ZGV mode will increase or decrease with changes in waveguide thickness, as the approximate wavelength of the $S_1S_2$-ZGV mode is given by $\lambda_{S_1S_2} \approx 3.7 \times 2h$~\cite{bruno_laser_2016}.

\begin{figure*}[ht!]
    \centering
    \includegraphics[width=0.7\linewidth]{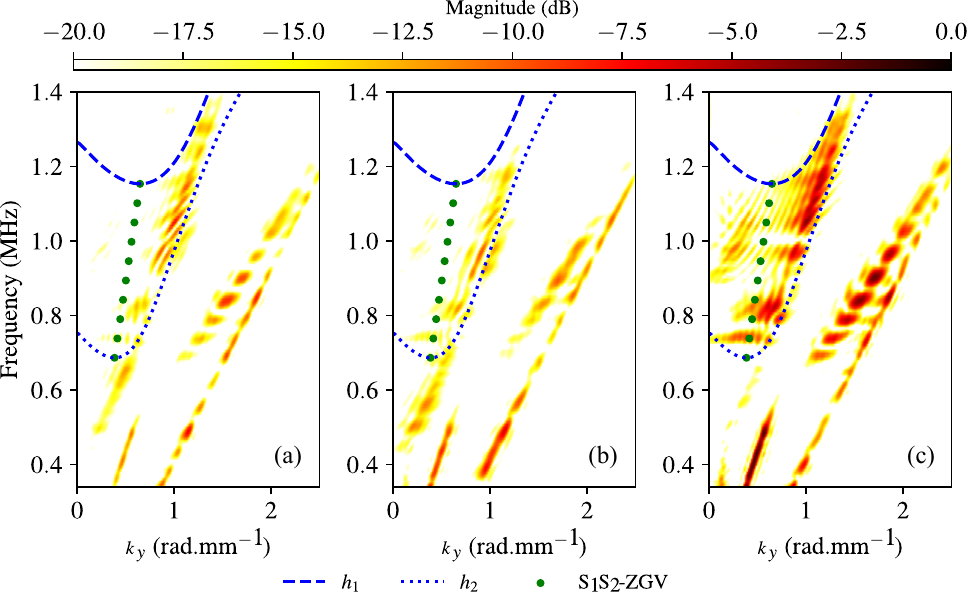}
    \caption{The experimental dispersion curves were obtained with L-transducer \#1 (a), \#2 (b), and the average of the two (c). The blue dots and dashes are the theoretical dispersion curves of the $S_1S_2$ branches. The green dots represent the locations of the $S_1S_2$-ZGV frequencies on the dispersion curve.}
    \label{average_disp}
\end{figure*}

To further investigate these phenomena, an additional transducer (\#2) was introduced to the experimental setup, positioned on the opposite side of the plate. This configuration enabled the averaging of the signals measured by the two transducers, suppressing antisymmetric modes, which exhibit out-of-phase displacements on opposite sides of the plate, while preserving symmetric modes, defined by in-phase displacements~\cite{auld_acoustic_1973} (see the Supplementary Material~\bluecolor{S6}). A comparison of Fig.~(\ref{average_disp}a) and Fig.~(\ref{average_disp}b) with Fig.~(\ref{average_disp}c), reveals that the application of summation enhances the visibility of symmetric modes, thereby facilitating the reduction of antisymmetric modes, in particular the crossover with $A_1$-mode~\cite{sohn2009, glushkov2010}. 

This represents a limitation of the thickness that can be reconstructed. It is noteworthy that the $S_0$-mode becomes more prominent, and the the S$_1$-mode, which is located within the band delimited by $h_1$ and $h_2$ is more distinctly observed. This configuration enables the averaging of the signals measured by the two transducers, thereby reducing the contribution of antisymmetric modes by cancelling out their out-of-phase displacements on opposite sides of the plate, while preserving symmetric modes that exhibit in-phase displacements. Nevertheless, the persistence of the $A_0$- and $A_1$-modes, despite their reduced intensity, suggests that antisymmetric modes have not been entirely suppressed. It is probable that this partial suppression is due to uncertainties in the precise alignment and couling of the transducers \#1 and \#2. However, the method is effective overall, as it enhances the visibility of symmetric modes, with the $S_0$-mode becoming more prominent and the $S_1$-mode in the band delimited by $h_1$ and $h_2$ appearing more distinctly.

Furthermore, this approach offers a distinctive advantage in plates with significant thickness variations, where higher-order modes can overlap at the same frequency. To illustrate, the $A_1$-mode and the $S_1S_2$-ZGV mode may coexist at the same frequency but manifest in disparate regions of the plate due to the thickness gradient. By isolating symmetric contributions, this method streamlines the detection and spatial localization of the $S_1S_2$-ZGV mode, rendering it especially advantageous in intricate waveguide geometries. In order to ascertain the $S_1S_2$-ZGV mode thicknesses, a comparable methodology is used, rise upon the processing techniques employed for the extraction of the $A_1$-mode cut-off thicknesses. 
\begin{figure}[b!]
    \centering
    \includegraphics[width=0.7\linewidth]{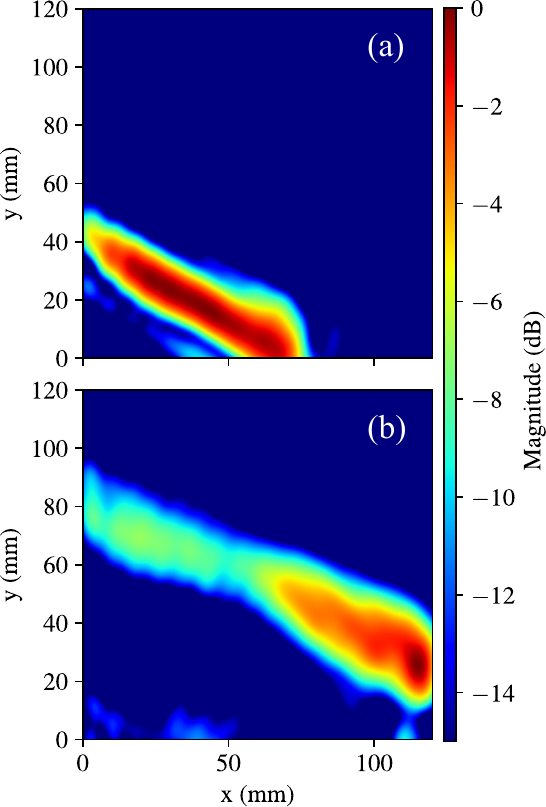}
    \caption{ZGV mode magnitude mapped at: (a) 0.76 MHz and (b) 0.9 MHz. We can observe the variation of the reconstructed ZGV bandwidth, which evolves with the thickness variation, as $\lambda_{S_1S_2} \approx 3.7 \times 2h$. We also note a decrease in amplitude, mainly due to viscoelastic attenuation (less than 1 dB/m in the MHz frequency range~\cite{laurent2014}).}
    \label{zgv_resonances}
\end{figure}
In this instance, the objective is to ascertain the resonance associated with the minimum wavenumber on the $S_1$-mode dispersion curve. A wavenumber filter is applied in the 3D-Fourier domain, in a manner analogous to that employed for the $A_1$-mode. In this case, the minimum and maximum wavenumbers, denoted by \(k_{\text{min}}\) and \(k_{\text{max}}\) are positioned symmetrically around the $S_1S_2$-ZGV wavenumber, with a margin of \(\pm 0.2\) rad.mm$^{-1}$ at each ZGV frequency. This ensures the precise isolation of the mode and effectively captures its stationary oscillatory behavior. Following the application of the filter, an inverse FFT is performed to reconstruct the wavefield associated with the $S_1S_2$-ZGV mode. Figures~(\ref{zgv_resonances}a) and (\ref{zgv_resonances}b) illustrate the absolute value of the filtered wavefields at 0.76 MHz and 0.9 MHz, respectively, and demonstrate the spatial distribution of the $S_1S_2$-ZGV resonance for thicknesses of 3.76 mm and 3.1 mm. As anticipated, the spatial distribution of the ZGV mode clearly follows a diagonal pattern, reflecting the thickness variation of the plate. As shown in Fig.~(\ref{zgv_resonances}b), the absolute value of the wavefield is non-uniform along the diagonal. This asymmetry arises because the $S_1$-mode propagates over a longer distance on the left side of the plate than on the right.

\begin{figure}[b!]
    \centering
    \includegraphics[width=0.7\linewidth]{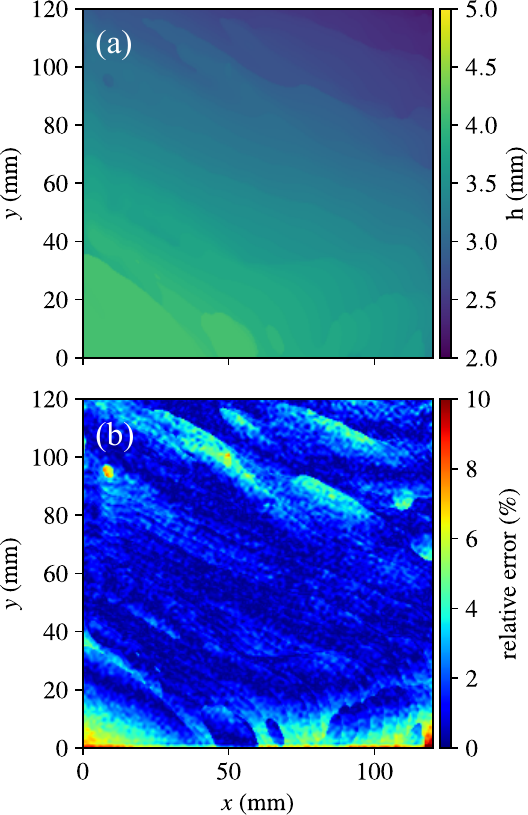}
    \caption{(a) Reconstructed thickness map and (b) Error between the thickness maps measured with the rangefinder and reconstructed using the ZGV-based method.}
    \label{final_res_ZGV}
\end{figure}

By applying this methodology at a range of $S_1S_2$-ZGV frequencies, a spatially resolved thickness map is produced. Figure~(\ref{final_res_ZGV}a) presents the reconstructed thickness map, which clearly reflects the expected thickness gradient. In comparison to the method utilizing the $A_1$-mode, the results demonstrate a notable enhancement.  

As shown in Fig.~(\ref{final_res_ZGV}b), the relative error map indicates a reduction in relative errors with this approach, achieving an average relative error of 1.43\%, which is approximately half that of the method based on the $A_1$-mode. This reduction in error can be attributed to several factors. In contrast to the $A_1$-mode, which operates near its cut-off frequency, the ZGV mode is distinguished by a fine local resonance. Furthermore, at the cut-off thickness for the $A_1$-mode, the associated wavenumber is close to zero, which can lead to energy leakage into evanescent modes. In contrast, the ZGV mode occurs at a thickness where the mode remains confined, with a finite wavenumber. This confinement allows for a more localized spatial distribution of the mode. As a result, the ZGV-based method provides a more precise representation of the thickness profile.

\section{Conclusion}\label{Conclusion}

In conclusion, this study has introduced and validated a comprehensive methodology for reconstructing the thicknesses of waveguides with varying thicknesses at all points of the plate. By leveraging the cut-off thickness of the $A_1$-mode and the $S_1S_2$-ZGV resonances remarkably induced at locations well beyond the laser source, we demonstrated the potential of combining adiabatic principles with experimental data to achieve accurate geometric reconstructions of elastic waveguides. The $A_1$-mode, which is sensitive to cut-off frequencies, is associated with a wavenumber equal to zero, thereby enabling the identification of cut-off thicknesses. In contrast, the $S_1S_2$-ZGV mode, which is characterized by its localized resonance and finite wavenumber, provides additional accuracy in regions with significant gradients. The ZGV-based method, in particular, offers reduction in errors and a more localized spatial distribution of energy, addressing limitations associated with the $A_1$-mode, such as energy leakage into evanescent modes near the cut-off thicknesses. 

The effectiveness of these approaches was validated by experimental results on additively manufactured metallic plates, which achieved a reconstructed thickness map with an average relative error of 1.43\% for the ZGV-thicknesses-based method. This is nearly twice as accurate as the $A_1$-mode approach. These methods have been proven effective for elastic waveguides with a thickness gradient. However, their versatility extends to waveguides with gradients in elasticity or temperature, making them invaluable for the precise characterization of inhomogeneous structures, including those observed in nature at considerably larger scales, but fully transferable, e.g. by monitoring the thicknesses of the pack of sea ice, and also extends to the field of seismology. 

Future work will focus on enhancing the selectivity of mode generation, aiming to eliminate the need for filtering in the 3D-Fourier domain. The use of a longitudinal transducer mounted on an adjustable wedge, or the integration of a continuously modulated laser, such as an EAM or VCSEL diode (optical valve), coupled with an optical Erbium amplifier in combination with a Spatial Light Modulator (SLM), could pave the way for a more efficient experimental setup to achieve this goal~\cite{grunsteidl_spatial_2013}. Ultimately, given the absence of an existing model describing the behavior of adiabatic Lamb modes at the cut-off or ZGV thicknesses, further analytical investigation in this domain would be highly beneficial.

\section*{AUTHOR DECLARATIONS}

\noindent\textbf{Acknowledgments.}
A.C. and J.L. are grateful for the funding provided by the European Commission under the European Union's Horizon 2020 research and innovation program (grant agreement 862017, \href{https://cordis.europa.eu/project/id/862017}{Grade2XL} EU project). The authors would like to thank K\'evyn Perlin for conducting the X-radiography of the AM-plate and for setting up the necessary equipment. They would also like to thank Karim Jezzine and Vahan Baronian for our thoughtful discussions and for implementing various upgrades on Scattering Matrix module available in \href{https://www.extende.com/fr/civa-en-quelques-mots}{\texttt{CIVA software}}.

\vspace{2mm}

\noindent\textbf{Author Contributions.}
The project was initiated by A.C. and J.L. J.L. and T.V-B. supervised the project. A.C. was responsible for coding the ultrasound acquisition sequences, conducting the experiments, and developing the post-processing tools. A.C., J.L., and T.V-B. analyzed the experimental results and theoretical study. A.C. was responsible for preparing the figures. A.C., J.L., and T.V-B. prepared the manuscript, discussing the results, and contributing to the finalization of the manuscript.

\vspace{2mm}

\noindent\textbf{Data availability.} The ultrasound data generated in this study are available from the corresponding author upon reasonable request.

\vspace{2mm}

\noindent\textbf{Code availability.}
Codes used to post-process the ultrasound data within this paper are available from the corresponding author upon reasonable request.\\




\begin{thebibliography}{0}%
\makeatletter
\providecommand \@ifxundefined [1]{%
 \@ifx{#1\undefined}
}%
\providecommand \@ifnum [1]{%
 \ifnum #1\expandafter \@firstoftwo
 \else \expandafter \@secondoftwo
 \fi
}%
\providecommand \@ifx [1]{%
 \ifx #1\expandafter \@firstoftwo
 \else \expandafter \@secondoftwo
 \fi
}%
\providecommand \natexlab [1]{#1}%
\providecommand \enquote  [1]{``#1''}%
\providecommand \bibnamefont  [1]{#1}%
\providecommand \bibfnamefont [1]{#1}%
\providecommand \citenamefont [1]{#1}%
\providecommand \href@noop [0]{\@secondoftwo}%
\providecommand \href [0]{\begingroup \@sanitize@url \@href}%
\providecommand \@href[1]{\@@startlink{#1}\@@href}%
\providecommand \@@href[1]{\endgroup#1\@@endlink}%
\providecommand \@sanitize@url [0]{\catcode `\\12\catcode `\$12\catcode
  `\&12\catcode `\#12\catcode `\^12\catcode `\_12\catcode `\%12\relax}%
\providecommand \@@startlink[1]{}%
\providecommand \@@endlink[0]{}%
\providecommand \url  [0]{\begingroup\@sanitize@url \@url }%
\providecommand \@url [1]{\endgroup\@href {#1}{\urlprefix }}%
\providecommand \urlprefix  [0]{URL }%
\providecommand \Eprint [0]{\href }%
\providecommand \doibase [0]{https://doi.org/}%
\providecommand \selectlanguage [0]{\@gobble}%
\providecommand \bibinfo  [0]{\@secondoftwo}%
\providecommand \bibfield  [0]{\@secondoftwo}%
\providecommand \translation [1]{[#1]}%
\providecommand \BibitemOpen [0]{}%
\providecommand \bibitemStop [0]{}%
\providecommand \bibitemNoStop [0]{.\EOS\space}%
\providecommand \EOS [0]{\spacefactor3000\relax}%
\providecommand \BibitemShut  [1]{\csname bibitem#1\endcsname}%
\let\auto@bib@innerbib\@empty
\end{thebibliography}%


%


\begin{thebibliography}{10}

\bibitem{pagneux_lamb_2006}
V.~Pagneux and A.~Maurel, ``Lamb wave propagation in elastic waveguides with
  variable thickness'',
  \emph{\href{http://dx.doi.org/10.1098/rspa.2005.1612}{Royal Soc. London Proc.
  Series A}} \textbf{462} (2006).

\bibitem{postnova_trapped_2007}
J.~Postnova and R.~V. Craster, ``Trapped modes in topographically varying
  elastic waveguides'', \emph{Wave Motion}  (2007).

\bibitem{Valier2008}
T.~Valier-Brasier, C.~Potel, and M.~Bruneau, ``Modes coupling of shear acoustic
  waves polarized along a one-dimensional corrugation on the surfaces of an
  isotropic solid plate'',
  \emph{\href{http://dx.doi.org/10.1063/1.2999632}{Appl. Phys. Lett.}}
  \textbf{93} (2008).

\bibitem{zima_theoretical_2022}
B.~Zima and J.~Moll, ``Theoretical and experimental analysis of guided wave
  propagation in plate-like structures with sinusoidal thickness variations'',
  \emph{\href{http://dx.doi.org/10.1007/s43452-022-00564-9}{Archives Civil
  Mech. Eng.}} \textbf{23}, 34 (2022).

\bibitem{biryukov1995}
S.~Biryukov, Y.~Gulyaev, V.~Krylov, and V.~Plessky,
  \href{http://dx.doi.org/10.1007/978-3-642-57767-3_9}{\emph{Rayleigh Waves on
  Curved Surfaces of Arbitrary Form. In: Surface Acoustic Waves in
  Inhomogeneous Media}}.
\newblock volume~20, Springer (1995).

\bibitem{lamb_waves_1917}
H.~Lamb, ``On waves in an elastic plate'',
  \emph{\href{http://dx.doi.org/10.1098/rspa.1917.0008}{Proc. Royal Soc.
  London. Series A}} \textbf{93}, 114--128 (1917), publisher: Royal Society.

\bibitem{pierce_extension_1965}
A.~Pierce, ``Extension of the {Method} of {Normal} {Modes} to {Sound}
  {Propagation} in an {Almost}-{Stratified} {Medium}'',
  \emph{\href{http://dx.doi.org/10.1121/1.1909303}{J. Acoust. Soc. Am.}}
  \textbf{37}, 19 (1965).
  
 \bibitem{santini_elastic_2023}
J.~Santini, E.~Riva, ``Elastic temporal waveguiding'',
  \emph{\href{http://dx.doi.org/10.1088/1367-2630/acb45d}{New J. Phys.}}
  \textbf{25}, 013031 (2023).

\bibitem{born_zur_1927}
M.~Born and R.~Oppenheimer, ``Zur {Quantentheorie} der {Molekeln}'',
  \emph{\href{http://dx.doi.org/10.1002/andp.19273892002}{Annalen der Physik}}
  \textbf{389}, 457--484 (1927).

\bibitem{born_beweis_1928}
M.~Born and V.~Fock, ``Beweis des {Adiabatensatzes}'',
  \emph{\href{http://dx.doi.org/10.1007/BF01343193}{Zeitschrift für Physik}}
  \textbf{51}, 165--180 (1928).

\bibitem{pierce_guided_1982}
A.~Pierce, ``Guided mode disappearance during upslope propagation in variable
  depth shallow water overlying a fluid bottom'',
  \emph{\href{http://dx.doi.org/10.1121/1.388033}{J. Acoust. Soc. Am.}}
  \textbf{72}, 523--531 (1982).

\bibitem{arnold_local_1986}
J.~M. Arnold and L.~B. Felsen, ``Local intrinsic modes: {Layer} with nonplanar
  interface'', \emph{\href{http://dx.doi.org/10.1016/0165-2125(86)90002-8}{Wave
  Motion}} \textbf{8}, 1--14 (1986).

\bibitem{el-kettani_guided_2004}
M.~E.-C. El-Kettani, F.~Luppé, and A.~Guillet, ``Guided waves in a plate with
  linearly varying thickness: experimental and numerical results'',
  \emph{\href{http://dx.doi.org/10.1016/j.ultras.2004.01.071}{Ultrasonics}}
  \textbf{42}, 807--812 (2004).

\bibitem{marical_guided_2007}
P.~Marical, M.~E.-C. El-Kettani, and M.~Predoi, ``Guided waves in elastic
  plates with gaussian section variation: Experimental and numerical results'',
  \emph{\href{http://dx.doi.org/10.1016/j.ultras.2007.05.004}{Ultrasonics}}
  \textbf{47}, 1--9 (2007).

\bibitem{Nakamura2012}
N.~Nakamura, H.~Ogi, M.~Hirao, and K.~Nakahata, ``Mode conversion behavior of
  {SH} guided wave in a tapered plate'',
  \emph{\href{http://dx.doi.org/10.1016/j.ndteint.2011.10.004}{NDT\&E Int.}}
  \textbf{45}, 156--161 (2012).

\bibitem{Nakamura2013}
N.~Nakamura, H.~Ogi, and M.~Hirao, ``Mode conversion and total reflection of
  torsional waves for pipe inspection'',
  \emph{\href{http://dx.doi.org/10.7567/JJAP.52.07HC14}{Jap. J. Appl. Phys.}}
  \textbf{52}, 07HC14 (2013).

\bibitem{arnold_intrinsic_1984}
J.~M. Arnold and L.~B. Felsen, ``Intrinsic modes in a nonseparable ocean
  waveguide'', \emph{\href{http://dx.doi.org/10.1121/1.391309}{J. Acoust. Soc.
  Am.}} \textbf{76}, 850--860 (1984).

\bibitem{felsen_adiabatic_1991}
L.~Felsen and L.~Sevgi, ``Adiabatic and intrinsic modes for wave propagation in
  guiding environments with longitudinal and transverse variation: formulation
  and canonical test'', \emph{\href{http://dx.doi.org/10.1109/8.97347}{IEEE
  Trans. Ant. Propa.}} \textbf{39}, 1130--1136 (1991), conference Name: IEEE
  Transactions on Antennas and Propagation.

\bibitem{rose_using_1998}
J.~Rose and J.~Barshinger, ``Using ultrasonic guided wave mode cutoff for
  corrosion detection and classification'',
  \href{http://dx.doi.org/10.1109/ULTSYM.1998.762277}{in \emph{1998 {IEEE}
  Ultra. Symp. Proc.}}, volume~1, pages 851--854 vol.1 (1998), iSSN: 1051-0117.

\bibitem{belanger_high_2014}
P.~Belanger, ``High order shear horizontal modes for minimum remnant
  thickness'',
  \emph{\href{http://dx.doi.org/10.1016/j.ultras.2013.12.013}{Ultrasonics}}
  \textbf{54}, 1078--1087 (2014).

\bibitem{suresh_remnant_2022}
N.~Suresh and K.~Balasubramaniam, ``Remnant thickness quantification in small
  thickness structures utilising the cut-off property of {A1} {Lamb} wave mode
  employing linear array elements'',
  \emph{\href{http://dx.doi.org/10.1063/5.0085102}{J. Appl. Phys.}}
  \textbf{131}, 174502 (2022).

\bibitem{niclas_reconstruction_2023}
A.~Niclas and L.~Seppecher, ``Reconstruction of smooth shape defects in
  waveguides using locally resonant frequencies'',
  \emph{\href{http://dx.doi.org/10.1088/1361-6420/acc7c0}{Inverse Problems}}
  \textbf{39}, 055006 (2023), publisher: IOP Publishing.

\bibitem{hamitouche_reflection_2009}
Z.~Hamitouche, M.~E.-C. El-Kettani, J.-L. Izbicki, and H.~Djelouah,
  ``Reflection at the {Cut}-off and {Transmission} by {Tunnel} {Effect} in a
  {Waveguide} with {Linear} {Section} {Variation}'',
  \emph{\href{http://dx.doi.org/10.3813/AAA.918209}{Acta Acustica}}
  \textbf{95}, 789--794 (2009).

\bibitem{yan_conversion_2015}
X.~Yan and F.-G. Yuan, ``Conversion of evanescent {Lamb} waves into propagating
  waves via a narrow aperture edge'',
  \emph{\href{http://dx.doi.org/10.1121/1.4921599}{J. Acoust. Soc. Am.}}
  \textbf{137}, 3523--3533 (2015).

\bibitem{germano_tunnel_2024}
M.~Germano, ``Tunnel effect for ultrasonic waves in tapered waveguides'',
  \emph{\href{http://dx.doi.org/10.3390/acoustics6020019}{Acoustics}}
  \textbf{6}, 362--373 (2024).

\bibitem{prada_laser-based_2005}
C.~Prada, O.~Balogun, and T.~W. Murray, ``Laser-based ultrasonic generation and
  detection of zero-group velocity {Lamb} waves in thin plates'',
  \emph{\href{http://dx.doi.org/10.1063/1.2128063}{Appl. Phys. Lett.}}
  \textbf{87}, 194109 (2005).

\bibitem{ces_thin_2011}
M.~Cès, D.~Clorennec, D.~Royer, and C.~Prada, ``Thin layer {Characterization}
  by {ZGV} {Lamb} modes'',
  \emph{\href{http://dx.doi.org/10.1088/1742-6596/269/1/012017}{J. Phys.: Conf.
  Series}} \textbf{269}, 012017 (2011).

\bibitem{yan2018}
G.~Yan, S.~Raetz, N.~Chigarev, V.~E. Gusev, and V.~Tournat, ``Characterization
  of progressive fatigue damage in solid plates by laser ultrasonic monitoring
  of zero-group-velocity lamb modes'',
  \emph{\href{http://dx.doi.org/10.1103/PhysRevApplied.9.061001}{Phys. Rev.
  Appl.}} \textbf{9}, 061001 (2018).

\bibitem{legrand2018}
F.~Legrand, B.~G{\'e}rardin, J.~Laurent, C.~Prada, and A.~Aubry, ``Negative
  refraction of lamb modes: A theoretical study'',
  \emph{\href{http://dx.doi.org/10.1103/PhysRevB.98.214114}{Phy. Rev. B}}
  \textbf{98}, 214114 (2018).

\bibitem{kiefer2023}
D.~A. Kiefer, S.~Mezil, and C.~Prada, ``Beating resonance patterns and extreme
  power flux skewing in anisotropic elastic plates'',
  \emph{\href{http://dx.doi.org/10.1126/sciadv.adk6846}{Sci. Adv.}} \textbf{9},
  eadk6846 (2023).

\bibitem{thelen2021}
M.~Thelen, N.~Bochud, M.~Brinker, C.~Prada, and P.~Huber, ``Laser-excited
  elastic guided waves reveal the complex mechanics of nanoporous silicon'',
  \emph{\href{http://dx.doi.org/10.1038/s41467-021-23398-0}{Nat. Comm.}}
  \textbf{12}, 3597 (2021).

\bibitem{morales_acoustoelastic_2024}
R.~E. Morales, N.~Pathak, J.~S. Lum, C.~M. Kube, T.~W. Murray, and D.~M.
  Stobbe, ``Acoustoelastic characterization of plates using zero group velocity
  {Lamb} modes'', \emph{\href{http://dx.doi.org/10.1063/5.0183620}{Appl. Phys.
  Lett.}} \textbf{124}, 084101 (2024).

\bibitem{scruby_laser_1990}
C.~B. Scruby and L.~E. Drain, \emph{Laser {Ultrasonics} {Techniques} and
  {Applications}}.
\newblock CRC Press, Oxford (1990).

\bibitem{achenbach2004}
J.~D. Achenbach,
  \href{http://dx.doi.org/10.1017/CBO9780511550485}{\emph{Reciprocity in elastodynamics}}.
\newblock Cambridge University Press (2004).  

\bibitem{moreau_measuring_2014}
L.~Moreau, J.-G. Minonzio, M.~Talmant, and P.~Laugier, ``Measuring the
  wavenumber of guided modes in waveguides with linearly varying thickness'',
  \emph{\href{http://dx.doi.org/10.1121/1.4869691}{J. Acoust. Soc. Am.}}
  \textbf{135}, 2614--2624 (2014).

\bibitem{royer_ondes_2021}
D.~Royer and T.~Valier-Brasier, \emph{Ondes élastiques dans les solides 2}.
\newblock volume~2 of \emph{Collection {Ondes}}, ISTE éditions, London (2021).

\bibitem{clorennec_local_2007}
D.~Clorennec, C.~Prada, and D.~Royer, ``Local and noncontact measurements of
  bulk acoustic wave velocities in thin isotropic plates and shells using zero
  group velocity {Lamb} modes'',
  \emph{\href{http://dx.doi.org/10.1063/1.2434824}{J. Appl. Phys.}}
  \textbf{101}, 034908 (2007).

\bibitem{grunsteidl_determination_2018}
C.~Grünsteidl, T.~Berer, M.~Hettich, and I.~Veres, ``Determination of
  thickness and bulk sound velocities of isotropic plates using
  zero-group-velocity {Lamb} waves'',
  \emph{\href{http://dx.doi.org/10.1063/1.5034313}{Appl. Phys. Lett.}}
  \textbf{112}, 251905 (2018).

\bibitem{fraser_orthogonality_1976}
W.~B. Fraser, ``Orthogonality relation for the {Rayleigh}–{Lamb} modes of
  vibration of a plate'', \emph{\href{http://dx.doi.org/10.1121/1.380851}{J.
  Acoust. Soc. Am.}} \textbf{59}, 215--216 (1976).

\bibitem{royer_elastic_2022}
D.~Royer and T.~Valier-Brasier, \emph{Elastic {Waves} in {Solids}, {Volume} 1:
  {Propagation}}.
\newblock Wiley-ISTE, Hoboken, 1st edition edition (2022).

\bibitem{bruno_laser_2016}
F.~Bruno, J.~Laurent, P.~Jehanno, D.~Royer, and C.~Prada, ``Laser beam shaping
  for enhanced {Zero}-{Group} {Velocity} {Lamb} modes generation'',
  \emph{\href{http://dx.doi.org/10.1121/1.4965291}{J. Acoust. Soc. Am.}}
  \textbf{140}, 2829--2838 (2016).

\bibitem{auld_acoustic_1973}
B.~A. Auld, \href{http://dx.doi.org/10.1063/1.3128926}{\emph{Acoustic fields
  and waves in solids}}.
\newblock volume~2, John Wiley \& Som, Inc. (1973).

\bibitem{sohn2009}
H.~Sohn and S.~B. Kim, ``Development of dual pzt transducers for reference-free
  crack detection in thin plate structures'',
  \emph{\href{http://dx.doi.org/10.1109/TUFFC.2010.1401}{IEEE Trans. Ultra.
  Ferro. Freq. Control}} \textbf{57}, 229--240 (2009).

\bibitem{glushkov2010}
E.~Glushkov, N.~Glushkova, O.~Kvasha, and R.~Lammering, ``Selective {Lamb} mode
  excitation by piezoelectric coaxial ring actuators'',
  \emph{\href{http://dx.doi.org/10.1088/0964-1726/19/3/035018}{Smart Mat.
  Struct.}} \textbf{19}, 035018 (2010).

\bibitem{laurent2014}
J.~Laurent, D.~Royer, and C.~Prada, ``Temporal behavior of laser induced
  elastic plate resonances'',
  \emph{\href{http://dx.doi.org/10.1016/j.wavemoti.2014.04.001}{Wave Motion}}
  \textbf{51}, 1011--1020 (2014).

\bibitem{grunsteidl_spatial_2013}
C.~Grünsteidl, I.~A. Veres, J.~Roither, P.~Burgholzer, T.~W. Murray, and
  T.~Berer, ``Spatial and temporal frequency domain laser-ultrasound applied in
  the direct measurement of dispersion relations of surface acoustic waves'',
  \emph{\href{http://dx.doi.org/10.1063/1.4773234}{Appl. Phys. Lett.}}
  \textbf{102}, 011103 (2013).
  

\end{thebibliography}



\end{document}



\makeatletter
\let\ps@titlepage\ps@fancy
\makeatother

\preprint{Suppl. Mat \#Ultrasonics-2025}

\title{SUPPLEMENTARY MATERIAL\\ Adiabatic Lamb modes in 3D tapered waveguides: Cut-off effects and ZGV resonances}

\author{Alexandre Yoshitaka Charau}
\email{alexandre.charau@cea.fr}
\affiliation{\href{https://ror.org/03xjwb503}{Universit\'e Paris-Saclay}, \href{https://ror.org/000dbcc61}{CEA, List}, F-91120, Palaiseau, France}
\affiliation{\href{https://ror.org/043we9s22}{Institut Jean Le Rond $\partial$'Alembert}, \href{https://ror.org/02en5vm52}{Sorbonne Universit\'e}, UMR CNRS7190, 4 Place Jussieu, Paris 75005, France\vskip 3mm\footnotesize{(Received 26 February 2025; Received in revised form 28 May 2025; Accepted 14 June 2025)}}

\author{J\'er\^ome Laurent}
\email{jerome.laurent2@cea.fr}
\affiliation{\href{https://ror.org/03xjwb503}{Universit\'e Paris-Saclay}, \href{https://ror.org/000dbcc61}{CEA, List}, F-91120, Palaiseau, France}

\author{Tony Valier-Brasier}
\email{tony.valier-brasier@sorbonne-universite.fr}
\affiliation{\href{https://ror.org/043we9s22}{Institut Jean Le Rond $\partial$'Alembert}, \href{https://ror.org/02en5vm52}{Sorbonne Universit\'e}, UMR CNRS7190, 4 Place Jussieu, Paris 75005, France\vskip 3mm\footnotesize{(Received 26 February 2025; Received in revised form 28 May 2025; Accepted 14 June 2025)}}


\begin{abstract}
The supplementary material provides further information: (\textit{i}) An X-ray map of the 2D gradient plate produced by LPBF; (\textit{ii}) The calculation of the transmission (T) and reflection (R) coefficients (scattering matrix method), with particular emphasis on the use of adiabatic modes for the determination of thickness variation; (\textit{iii}) The excitability of the modes and the anticipated displacement field as a function of thickness variation; (\textit{iv}) Additionally, space-time associated diagrams are presented with superimposed theoretical times-of-flight, and the spectral spread of the modal energy overlaid on the theoretical dispersion curves. (\textit{v}) Determining wave velocities from the two first ZGV frequencies. (\textit{vi}) Selective detection of symmetrical modes.
\vskip 2mm
\noindent DOI: \href{https://ars.els-cdn.com/content/image/1-s2.0-S0041624X25001702-mmc1.pdf}{\nolinkurl{1-s2.0-S0041624X25001702-mmc1}}
\end{abstract}

\maketitle

\section{X-ray imaging of graded thickness plate}
\label{sm:xray_plate}

To ensure that the sample is free of defects such as cracks or microporosities, an X-ray inspection is performed using the Kuka robot (KR-C4 HA). The process uses the Viscom (X9225-DED) as an emitter and the XRD (1621 AN/CN) digital X-ray detector from Perkin Elmer. During the inspection, the sample and the detector remain stationary, while only the emitter moves. A 264$\times$264 mm$^2$ scan is performed with a step size of 16.5 mm, providing a spatial resolution of approximately 10 µm. 
\begin{figure*}[ht!]
\centering
    \includegraphics[width=0.45\linewidth]{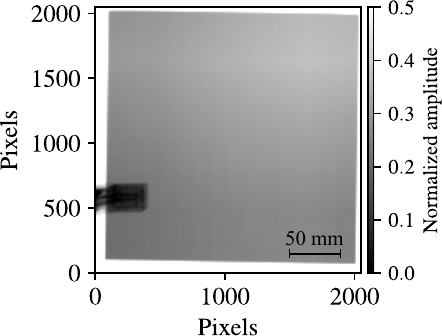}
    \caption{X-ray map of the AM-plate reconstructed from multiple scans with a ROI of 16.5 x 16.5 mm$^2$. No cracks or microporosities were observed. We also note the variation in gray level, which is directly proportional to the variation in thickness of the inhomogeneous waveguide.}
  \label{sm:Xray}
\end{figure*}
The result is shown in Fig.(~\ref{sm:Xray}). In addition to confirming the absence of defects, the grayscale map gradient shows the thickness variations in the sample, in accordance with the thickness measurement by means of the laser telemeter above. Here, higher values indicate better X-ray transmission.

\section{Determining wave velocities from ZGV frequencies}
\label{sm:ZGV_spectrum}

\begin{figure*}[ht!]
\centering
    \includegraphics[width=0.7\linewidth]{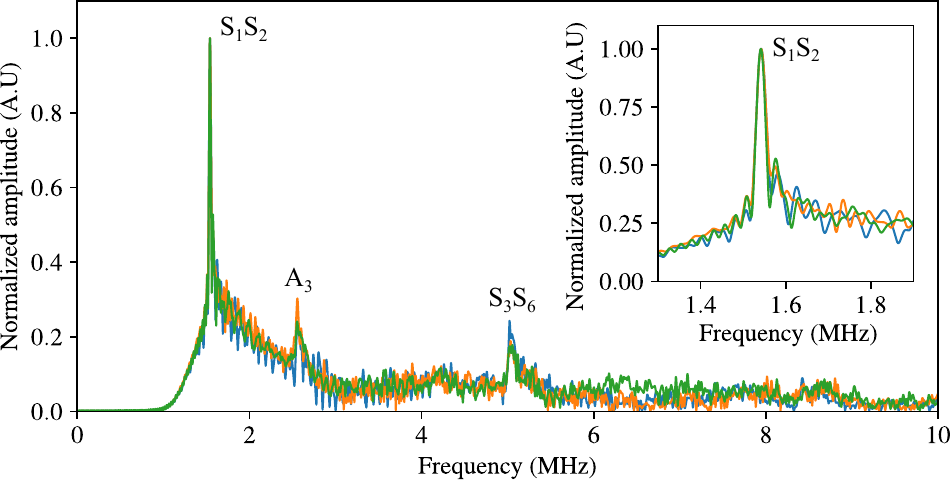}
    \caption{Spectrum of signals (normal displacement) measured at various points on an AM plate of constant thickness.}
  \label{sm:zgv_spectre}
\end{figure*}

Figure~\ref{sm:zgv_spectre} illustrates the spectra obtained from measurements performed with the Quartet interferometer and a Q-switched Nd:YAG (Quantum Light, Q2HE) pulsed laser at various points on the AM plate with a constant thickness of 1.85 mm. The results demonstrate unequivocally that the $S_1S_2$-ZGV mode is established at the same frequency, thereby underscoring the repeatability and reliability of the experimental procedure. By analyzing the average spectrum, the frequency of the $f_{S_1S_2}$ mode was determined to be 1.54 MHz, which enabled the extraction of the longitudinal and transverse wave velocities ($V_L$ and $V_T$). The resulting bulk wave velocities are: V$_L$ = 6.32 µs.mm$^{-1}$ and V$_T$ = 3.12 µs.mm$^{-1}$ ($\nu = 0.34$).

\section{Scattering Matrix: reflection (R) and transmission (T) coefficients}
\label{sm:Adiabadicity}

\begin{figure}[ht!]
\centering
    \includegraphics[width=0.8\linewidth]{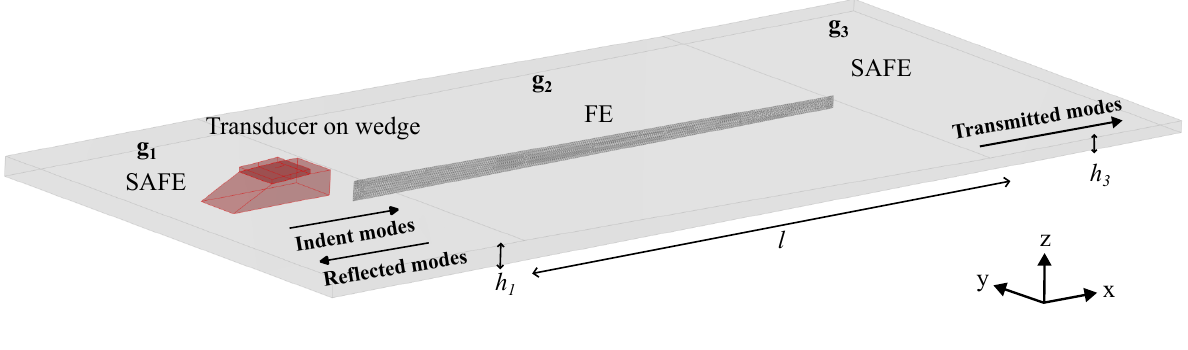}
    \caption{Civa simulation configuration: A$_1$-mode excited with a 3.9° incident angle wedge in a waveguide with linear thickness variation. g$_1$ and g$_3$ are the waveguides in which the SAFE calculation is performed, g$_2$ the meshed waveguide in which the spectral FE calculation is achieved to obtain the energy of incident (I), transmitted (T), and diffracted (D) modes .}
  \label{sm:conf_civa}
\end{figure}

The objective of this section is to illustrate the behavior of the A$_1$-mode in an inhomogeneous guide with a linear thickness variation and to demonstrate that it adheres to the adiabatic condition. To observe the behavior in question, a simulation was conducted using the Guided Waves module of the CIVA software~\citep{baronian_hybrid_2011} based on the scattering matrix method~\citep{varatharajulu1976, pagneux2004}. The configuration, depicted in Fig.~(\ref{sm:conf_civa}), comprises an aluminium plate partitioned into three sections: the initial section, g$_1$ has a thickness $h_1$ of 4.2 mm, while the final section g$_3$ has a waveguide thickness $h_3$ of 2.5 mm. This results in the formation of a waveguide g$_2$ with a linear thickness variation $h_2(x)$ between $h_1$ and $h_3$ over a length $l$ of 120 mm: the spectral finite element area.

In contrast to the case study presented subsequently, the waveguide varies in only one direction, thereby facilitating a more straightforward comprehension of the behavior of A$_1$-mode. In order to facilitate the propagation of A$_1$-mode and ensure its attainment of the designated cut-off thickness ($h_c$) at 3.5 mm, a wedge is employed with an incidence angle of 3.9°, centered at 520 kHz. The calculation of the incidence angle is based on Snell's law.
\begin{figure*}[ht!]
\centering
    \includegraphics[width=0.7\linewidth]{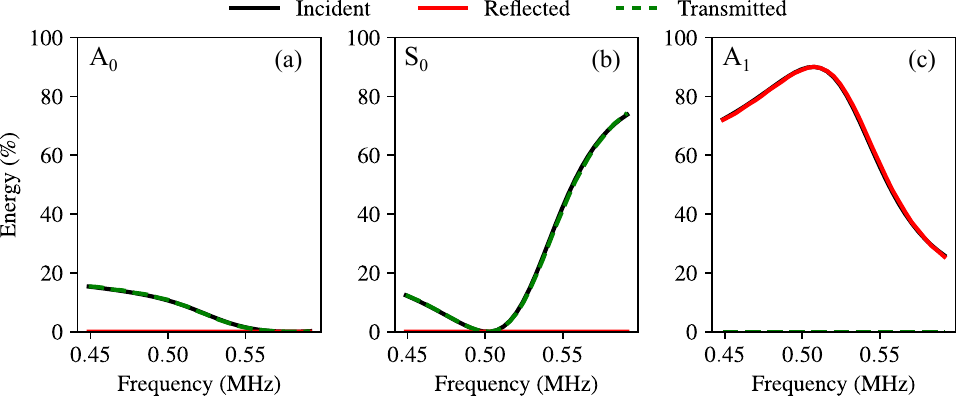}
    \caption{Energy of the incident (I), transmitted (T), or diffracted (D) modes. The energy conservation is satisfied, since $\sum\left|I_m^2+T_m^2+D_m^2 \right| \approx 1$.}
  \label{sm:IRT}
\end{figure*}

Figure~\ref{sm:IRT} shows the incidence, reflection, and transmission curves for modes: A$_0$ (a), S$_0$ (b), and A$_1$ (c). The incident modes are those emitted in guide g$_1$, the reflected modes are those that return to g$_1$, and the transmitted modes are those that pass through guide g$_2$ to reach g$_3$. In contrast to the A$_1$-mode, it is apparent that the fundamental modes are neither reflected nor diffracted in waveguide g$_2$. This phenomenon occurs because, for adiabatic modes, the fundamental modes lack cut-off frequencies. Consequently, they never encounter a cut-off thickness $h_2$, allowing them to maintain continuous phase matching throughout their propagation. Thus, it can be stated that in this guide, the reflection coefficients of the fundamental modes are zero, while for the A$_1$-mode, it is equal to one.

\section{Modal amplitudes: excitability}
\label{sm:excitability}

\begin{figure*}[ht!]
\centering
    \includegraphics[width=0.9\linewidth]{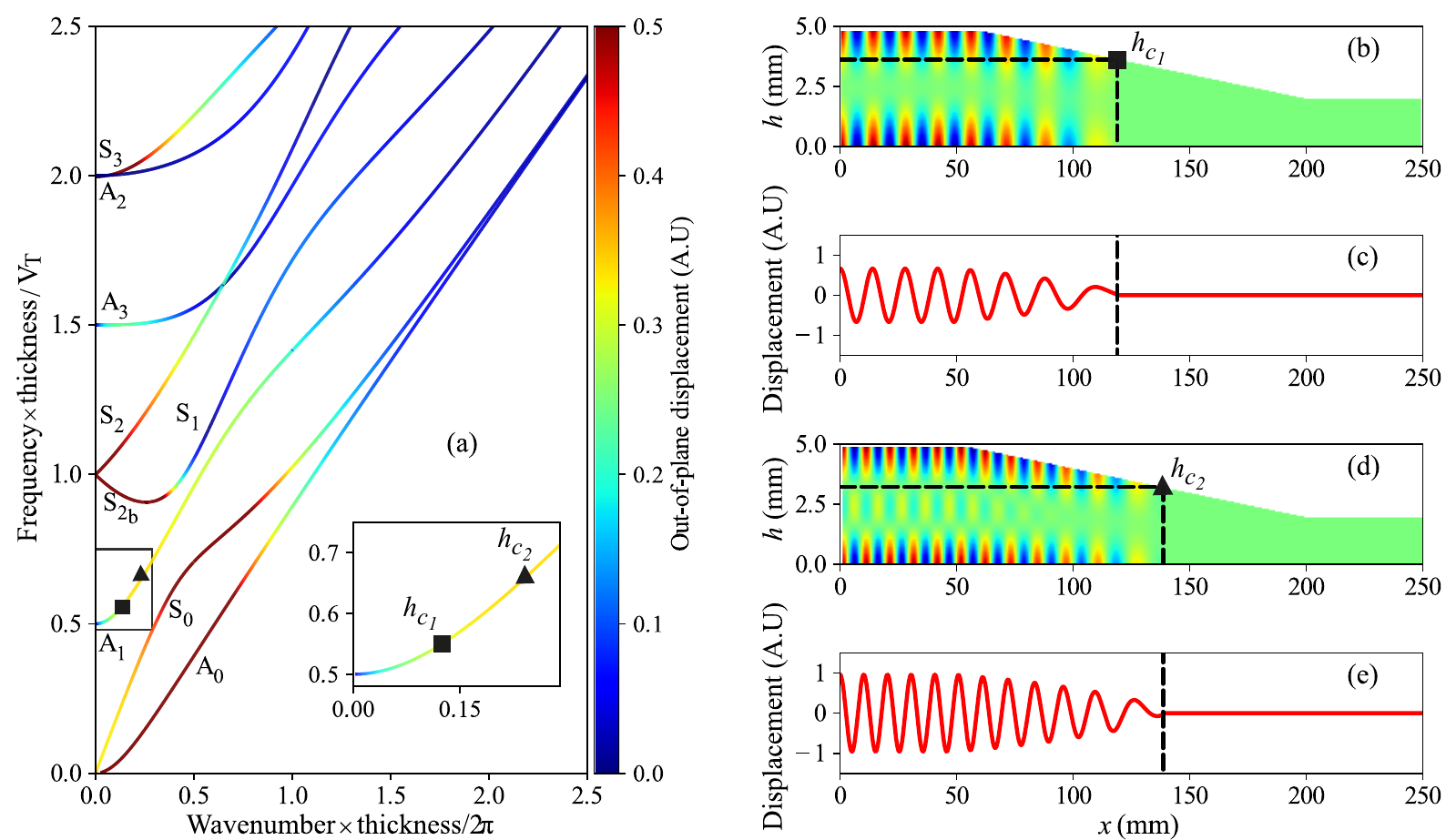}
    \caption{Theoretical laser excitability overlaid on the dimensionless dispersion curves for a 3 mm thick aluminum plate is shown in (a). Panels (b) through (e) display the normal displacement wavefield of the incident A$_1$-mode at different frequencies: 0.5 and 0.6 MHz. The cut-off thicknesses for the A$_1$-mode at these frequencies are denoted as $h_{c_1}$ and $h_{c_2}$, respectively.}
  \label{sm:champs_ana}
\end{figure*}

Figure~(\ref{sm:champs_ana}a) illustrates the excitability of the out-of-plane displacement of modes generated by a 7 ns pulsed laser with a 1 mm focal spot, superimposed on the dimensionless dispersion curves for an aluminum plate with a thickness of 3 mm. The theoretical values were calculated using a semi-analytical Python code that resolves the Rayleigh-Lamb equation to determine the dispersion curves. The excitability is established using the bi-orthogonality relations developed by Fraser~\cite{fraser_orthogonality_1976}, and later generalized by Gunawan~\cite{gunawan_reflection_2007}. As observed in Fig.~(\ref{sm:champs_ana}a), the out-of-plane displacement of the A$_1$-mode tends to zero at its cut-off frequency, in contrast to the behavior observed in other modes. This phenomenon can be attributed to the fact that the cut-off frequency of the A$_1$-mode, which is given by the expression $f_{c_{A_1}} = V_T/2h$ where $V_T$ represents the transverse velocity and $h$ is the thickness of the waveguide, is characterized by the propagation of a transverse bulk wave through the thickness, which lacks any out-of-plane displacement component.

\section{Modal amplitude and spectral energy distribution}
\label{sm:ModalAmplitude}

\begin{figure*}[ht!]
\centering
    \includegraphics[width=0.85\linewidth]{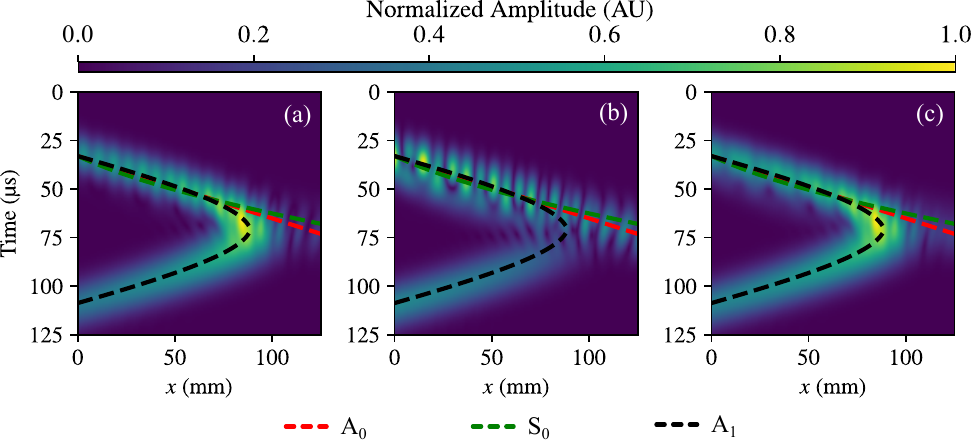}
    \caption{Envelop of the spatio-temporal diagram at $z=0$ mm of: in-plane displacement $u_x$ (a); out-of-plane displacement $u_z$ (b); and the absolute value of the displacement $|u|$.}
  \label{sm:bscan}
\end{figure*}

In order to gain a deeper insight into the transition between the A$_1$-mode and its cut-off thickness, signals were extracted at each point of the waveguide g$_2$ at $h=0$ mm. This enabled the observation of the spatio-temporal diagram. Figure~(\ref{sm:bscan}a) illustrates that A$_1$-mode remains undiffracted throughout its propagation until it reaches its cut-off thickness, indicating that the adiabatic condition is well maintained. Figure~(\ref{sm:bscan}b) does not display a reflection at the cut-off thickness due to the fact that, as previously outlined in reference~\ref{sm:Adiabadicity}, the normal component of the modal amplitude for the A$_1$-mode tends to zero at the cut-off frequency. Furthermore, the theoretical time-of-flight of incident modes is calculated using the following formula and superimposed on Fig.~(\ref{sm:bscan}):
\begin{equation*}
    t_{m}(x,f) = \sum_{n = 0}^{N-1}\frac{x_{n+1}-x_n}{V_{g{_m}}(h(x_n),f)},
\end{equation*}
where the variable $m$ represents the number, which can be attributed either A$_m$- or S$_m$-mode. The variable $N$ denotes the number of discretisation points along the waveguide until the cut-off thickness $h_c$, $V_{g{_m}}$ the group velocity of the mode and $f$ the frequency.

For higher-order modes, this formula requires a specific condition. When the mode approaches its cut-off thickness $h_c$, $V_{g{_m}}$ tends to zero. To prevent analytical singularities, we consider that when $V_{g{_m}}$ is on the order of 10$^{-3}$ mm/µs, the mode has effectively reached its cut-off thickness and ceases to propagate. The application of this formula, in conjunction with the notable concordance observed, serves to substantiate the assertion that the mode undergoes instantaneous and local adaptation in accordance with the section of the guide through which it propagates. This corroborates the existence of the adiabatic phenomenon, as postulated by A. Pierce \cite{pierce_extension_1965}. An alternative methodology for the observation of these phenomena is through the utilization of dispersion curves. 

\begin{figure*}[ht!]
\centering
    \includegraphics[width=0.75\linewidth]{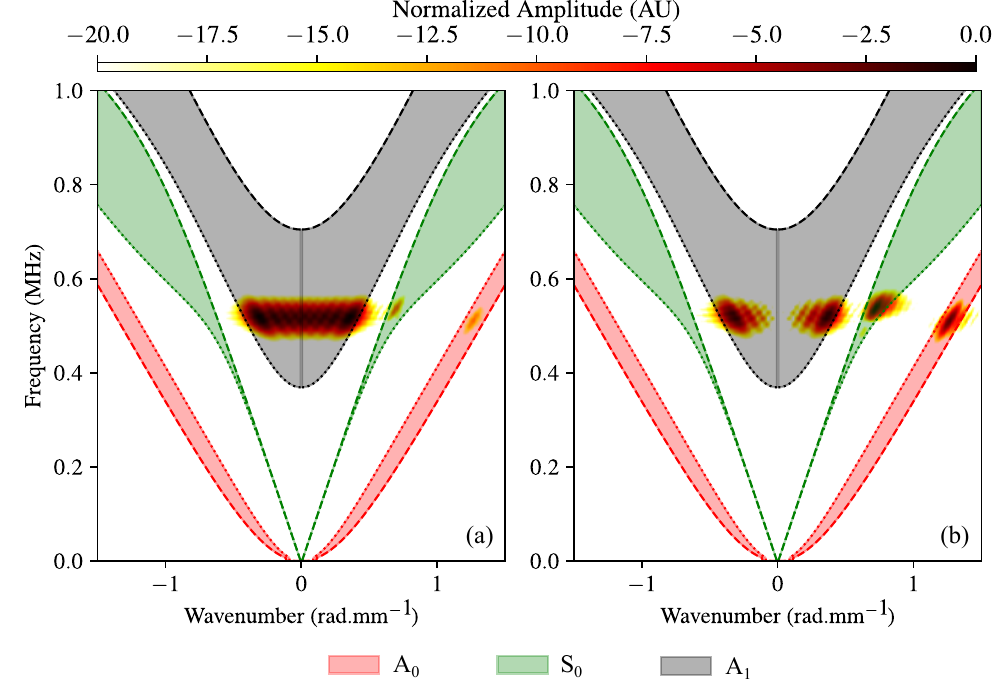}
    \caption{Theoretical dispersion curves superimposed on those obtained from numerical data: in-plane component (a); out-of-plane component (b); with (\dotted) for $h_1$ = 4.2 mm, and (\dashed) for $h_2$ = 2.5 mm.}
  \label{sm:CD}
\end{figure*}

Figure~(\ref{sm:CD}) presents theoretical dispersion curves superimposed on numerical data, demonstrating a high degree of concordance between the theoretical and empirical results. This representation elucidates the modal amplitude and energy distribution across the associated wavenumber band. As previously noted, the energy of the A$_1$-mode at its cut-off frequency is low for the normal component, and very high for the in-plane component. In the negative wavenumbers, only the A$_1$-mode is present, indicating that reflection occurs during propagation. Finally, the energy distribution across the wavenumber band clearly illustrates the adiabatic behavior of A$_1$-modes considered in this paper. These observations, not to mention the appearance of lower adiabatic modes, are almost transferable to higher cut-off frequencies and will be discussed in a later paper.

\section{Selective detection of symmetrical modes}
\label{sm:ZGV_spectrum}

\begin{figure}[ht!]
\centering
    \includegraphics[width=0.8\linewidth]{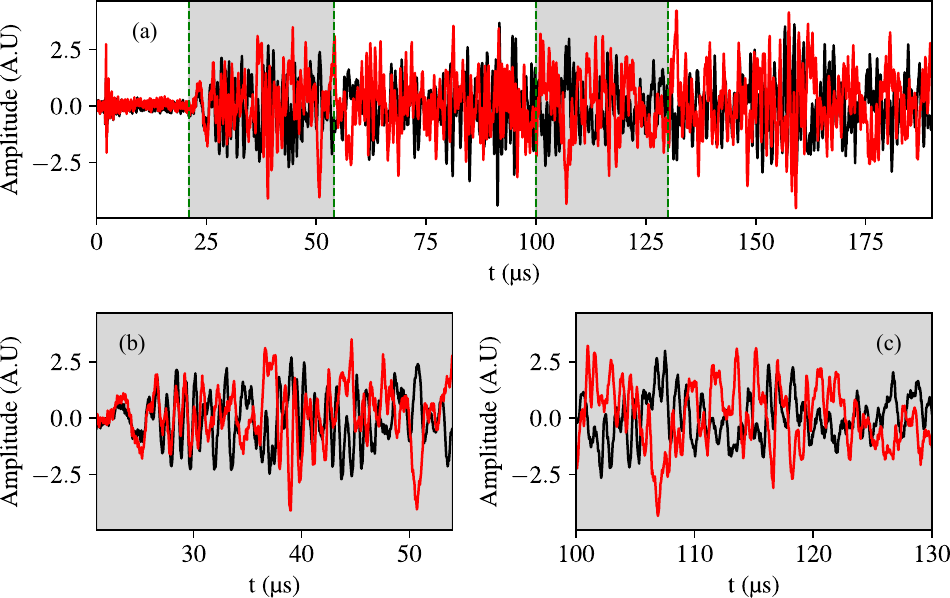}
    \caption{Signals of transducers \#1 (black) and \#2 (red) at $x = 60$~mm and $y = 60$~mm. Total measured signal (a) and zoomed signal at different instance (b and c).}
  \label{sm:sig12}
\end{figure}

To facilitate the propagation of a specific mode~\cite{philippe2015, serey2021} or mode family (A or S)~\citep{borkowski2013, zhang2023}, various techniques are available, most of which rely on selective generation methodologies. In this study, a pulsed laser is employed to generate all potential A- and S-modes across a broad bandwidth. To probe S-modes selectively, we applied a detection method based on the reciprocity principle. Figure~\ref{sm:sig12} illustrates the ultrasonic signals recorded by transducers \#1 and \#2, corresponding to a scenario where the laser source is positioned at the center of the scanning area (60, 60) mm.

A comparison of the signals reveals both in-phase and opposite-phase alignment. In-phase alignment indicates contributions from S-modes, whereas opposite-phase alignment reflects contributions from A-modes. By summing the signals from the two transducers, the S-mode contributions are enhanced, while the A-modes are significantly diminished. Conversely, subtracting the signals achieves the opposite effect. Nevertheless, the removal of A-modes remains incomplete. This limitation can be attributed to three main factors: variations in the transducers' impulse responses, coupling quality, and imprecise alignment of the transducers on each side of the AM-plate.

